\documentclass[proceedings]{JHEP} 

\usepackage{epsfig}			

\newbox\mybox
\newcommand\fverb{\setbox\mybox=\hbox\bgroup\verb}
\newcommand\fverbdo{\egroup\medskip\noindent\fbox{\unhbox\mybox}\ }
\newcommand\fverbit{\egroup\item[\fbox{\unhbox\mybox}]}

\font\beeg=cmr17 scaled 1600		
\newcommand\init[1]{\setbox\mybox=\hbox{{\beeg #1}~}%
		   \noindent\global\hangindent=\wd\mybox\global\hangafter-2%
		   \sc\smash{\llap {\lower 13.2pt \box\mybox}}}

\def\Rea{\rlap{\rm I}\mkern3mu{\rm R}}

\title{From ADM to Brane-World charges}

\author{Sebasti{\'a}n Silva \\

	Max Planck Institut f{\"u}r Gravitationsphysik, Albert Einstein
Institut, \\
Am M{\"u}hlenberg 5, D-14476 Golm, Germany\\
	E-mail: \email{silva@aei-potsdam.mpg.de}}

\conference{Nonperturbative Quantum Effects 2000}

\abstract{
We first recall a covariant formalism used to compute conserved
charges in gauge invariant theories. We then study the case of
gravity for two different boundary conditions, namely spatial
infinity and a Brane-World boundary.
The new conclusion of this analysis is that the gravitational energy
(and linear 
and angular momentum) is a {\it local} expression if our universe is really
a boundary of a five-dimensional spacetime.}

\begin{document} 

\maketitle 


\section{Introduction}\label{intm}

Recently,
the old idea that our real world could be a boundary of an higher
dimensional spacetime has been intensively revisited after the work
of Randall and Sundrum \cite{RS}.
This so-called Brane-World scenario has highly non-trivial consequences
on the effective four dimensional theory of gravity. In fact,
the equations of motion on the brane-universe are not the usual Einstein 
equations in four dimensions but the modified version given in
\cite{SMS}. 
Another big difference 
concerns the conserved charges. 

In ordinary four dimensional Einstein gravity, there is not such a
notion of a local density of energy in the bulk spacetime. 
This is a direct consequence of the local diffeomorphism invariance.
For an asymptotically flat spacetime, we can however construct a local
energy density {\it at spatial infinity}, to be integrated on ${\cal
B}_{\infty}$, see figure \ref{spacetime} (with $D=4$). This is the well known ADM
mass whose construction crucially depends on the asymptotically flat
boundary conditions \cite{RT}.


\smallskip
\FIGURE{\epsfig{file=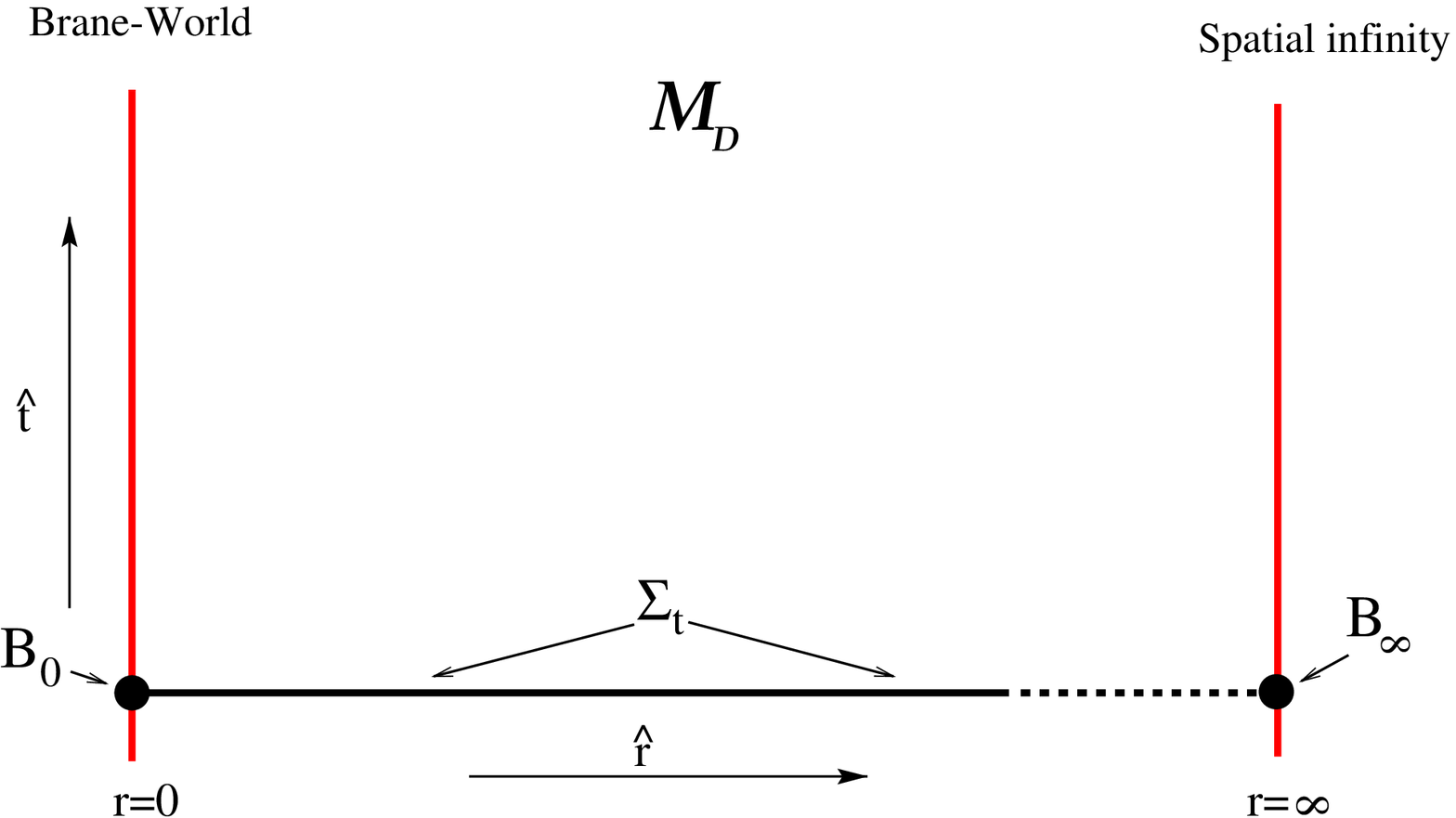,width=7.3cm}%
        \caption{Spacetime bounded by one Brane-World at $r=0$, by
spatial infinity at $r=\infty$ and by one Cauchy hypersurface $\Sigma _{t}$.}%
	\label{spacetime}}

Let us now consider a five-dimensional spacetime 
bounded by one D3-brane a $r=0$ (see again figure \ref{spacetime} but
now with $D=5$). 
Following the Brane-World scenario, ${\cal B}_{0}$ has
to be seen as a Cauchy hypersurface of our four-dimensional universe.
Now, we can
of course still define an ADM  
mass at ${\cal B}_{\infty}$ if the five-dimensional spacetime is
asymptotically flat. However, this mass cannot be measured since, by
hypothesis, we are not allowed to escape from the Brane-World.

On the other hand, we can construct a measurable 
local density of mass at ${\cal B}_{0}$ (similar to the ADM mass at
${\cal B}_{\infty}$) using now the boundary conditions
on the Brane (namely the Lanczos-Israel junction conditions). 
Then, although the gravitational energy is still non-local in
five dimensions, we have local expressions at {\it each} boundary
 of this higher dimensional spacetime (namely at  ${\cal B}_{0}, {\cal
B}_{\infty},$ or others if any). Therefore, the effective
four-dimensional gravitational energy becomes {\it localizable} within
${\cal B}_{0}$ in a Brane-World scenario.

In sections \ref{consch} and \ref{compsup} we recall the basic
construction of conserved charges in gauge invariant theories
following the references \cite{JS1,Si,SiT}. We then study in section
\ref{gravex}
the case of gravity in $D$ dimensions, for the two different
boundary conditions shown in figure \ref{spacetime}. For
asymptotically flat boundary conditions (namely Dirichlet boundary
conditions on the metric), we recover the ADM charges, in a generally
covariant form (the so-called KBL superpotential \cite{KBL})
\cite{JS3}. For the 
Lanczos-Israel boundary conditions \cite{LI} (of Neumann type on the
metric), we 
find a new covariant expression for the Brane-World charges \cite{Si2}. We
finally use this expression in two very simple examples.

We skip all the details in calculations which can be found in the
cited references.

\section{Conserved charges associated with gauge symmetries}\label{consch}

Let us assume that one theory is invariant under some symmetry
$\delta_{\xi } \varphi^{i}$. The label $^{i}$ goes for all the fields
present in the Lagrangian, even the auxiliary ones. 

For a {\it global} symmetry,
the infinitesimal
parameter $\xi^{\alpha }$ is {\it
constant}. The conserved charges are then given by the usual Noether construction: 
\begin{equation}\label{noechar}
Q_{\xi } = \int_{\Sigma_{t}} J^{\mu }_{\xi} \hat{t}_{\mu}
\end{equation}
where $J^{\mu }_{\xi}$ denotes the associated Noether current and
$\hat{t}_{\mu}$ the normal to the Cauchy hypersurface $\Sigma_{t}$. 

For a 
{\it local} or gauge symmetry, (that is,
$\xi^{\alpha}=\xi^{\alpha} (x)$), the formula (\ref{noechar}) is
modified by:

\begin{equation}\label{gauco}
Q_{\xi } = \int_{{\cal{B}}_{r}} U^{\mu \nu}_{\xi} \hat{t}_{\mu}\hat{r}_{\nu} 
\end{equation}
for $r=0$ {\it or} $\infty$; see figure \ref{spacetime}. We also
denoted by $\hat{t}_{\mu}$ and $\hat{r}_{\nu}$ the normal vectors of
${\cal{B}}_{r}$.

The tensor
$U^{\mu \nu}_{\xi}=-U^{\nu \mu}_{\xi}$ is called {\it
superpotential} and 
therefore plays the role of the Noether
current for local symmetries. Its construction will be recalled  
in the next section. 

The important points about equation (\ref{gauco}) are the
following. The  
conserved charges associated with a gauge symmetry \cite{JS1,Si,SiT,JS4}:

$\bullet$  can be computed on {\it each} boundary of spacetime
${\cal{B}}_{r}$ (for $r=0, \infty,$ {\it or others if any}; see figure
\ref{spacetime}) in a completely independent way by the formula 
(\ref{gauco});

$\bullet$ strongly depend on the boundary conditions imposed
on ${\cal{B}}_{r}$;

$\bullet$ are functional of the gauge parameter
$\xi^{\alpha} (x)$. Moreover, the number of conserved charges is
given by the number of ``non-zero''\footnote{We call
``non-zero'' (on ${\cal B}_{r}$) a gauge symmetry 
whose parameter $\xi^{\alpha}$ does not vanish on ${\cal B}_{r}$.} gauge symmetries $\delta_{\xi}
\varphi^{i}$ 
compatibles with the boundary
conditions on ${\cal B}_{r}$. 


\section{Computing the superpotential}\label{compsup}

The method presented in \cite{Si} to compute the superpotential
$U^{\mu \nu}_{\xi}$ (\ref{gauco}) can be summarized by the following
``recipe'':

i) The theory should be reformulated in a {\it first order
form}:
\begin{eqnarray}
\frac{\delta {\cal L}}{\delta \varphi^{i}} &=:& E_{i}  =
E_{i}(\varphi, \partial_{\mu} \varphi)\label{equ1} \\
\delta _{\xi} \varphi^{i} &=& \partial_{\mu} \xi^{\alpha} \Delta ^{\mu
i}_{\alpha } (\varphi, \partial_{\mu}
\varphi) +\xi ^{\alpha } \Delta ^{i}_{\alpha }(\varphi, \partial_{\mu}
\varphi) \ \ \ \ \ \label{ghj}
\end{eqnarray}

\noindent That is, both $E_{i}$, $\Delta ^{\mu
i}_{\alpha }$ and  $\Delta ^{i}_{\alpha }$ should not depend on
$\partial^{2} \varphi^{i}$, $\partial^{3} \varphi^{i}$, etc\dots The
introduction of auxiliary fields (which are included in the $^{i}$
index) is usually needed. 
For simply \cite{Si}, we also allow at
most one derivative of the gauge parameter
(that is, no terms in $\partial^{2} \xi $, $\partial^{3} \xi $,
etc\dots) in the symmetry transformation laws (\ref{ghj}).

ii) With the definitions (\ref{equ1}) and (\ref{ghj}), we construct the following tensor,
\begin{equation}\label{wdef}
W^{\mu }_{\xi } := \xi^{\alpha} \Delta ^{\mu i}_{\alpha } E_{i},
\end{equation}
which vanishes on-shell. Note that this tensor also appears in the {\it Noether
identities} associated with (\ref{ghj}), namely 
$\delta _{\xi} \varphi^{i} E_{i} = \partial _{\mu} W^{\mu }_{\xi }$.

iii) Then, an {\it arbitrary} variation of the
superpotential, on {\it any} ${\cal B}_{r}$, should satisfy:
\begin{equation}\label{varu}
\left. \delta U^{\mu \nu }_{\xi} = -\delta \varphi^{i} \frac{\partial
W_{\xi}^{\mu }}{\partial \partial_{\nu} \varphi^{i}} \right|_{on\  {\cal
B}_{r}}.
\end{equation}
A theorem \cite{Si} guarantees the antisymmetry in
$\mu $ and $\nu $ of the rhs of equation (\ref{varu}). 

iv) The last step is to ``integrate'' the equation
(\ref{varu}). That is, to rewrite the rhs of
(\ref{varu}) as $\delta (\mbox{\it something})$, using the boundary
conditions on ${\cal B}_{r}$.

\smallskip

The formula (\ref{varu}) gives an unambiguous superpotential, up to
some global ``integration''-constant. Moreover, it follows from 
three points:

\begin{enumerate}
\item [] $\bullet$ The Noether current associated with a gauge symmetry
can be rewritten as: 
\begin{equation}\label{noeth}
J_{\xi}^{\mu } = W^{\mu }_{\xi} + \partial_{\nu} U_{\xi}^{\mu \nu },
\end{equation}
for some $U_{\xi}^{\mu \nu }=-U_{\xi}^{\nu \mu }$ (together with the
definition (\ref{wdef})).
\item [] $\bullet$ The variational principle is satisfied on ${\cal B}_{r}$.
\item [] $\bullet$ The charge defined by (\ref{gauco}) is on-shell conserved.
\end{enumerate}

The first point can be proven in general \cite{JS1}. It
follows from the {\it locality} of the gauge symmetry considered. The second
point has to be checked by hand and then imposes some restrictions on the
``allowed'' boundary conditions: they should be compatible with some
variational principle on ${\cal B}_{r}$. 
The third point is required also by hand. Therefore,
the charge (\ref{gauco}) will be conserved if the
basic equation (\ref{varu}) for the superpotential holds.
The complete derivation of (\ref{varu}) can be found in \cite{Si,SiT,JS4}.

The method summarized by the four above steps i)-iv) is nothing but a
Lagrangian (and then explicitely covariant) version of the
Regge and Teitelboim procedure \cite{RT} in Hamiltonian formalism. There exists in
fact a precise correspondence between both methods, through the
so-called covariant symplectic phase space formalism \cite{WC,ABR}. The key point is
to realize that the equation (\ref{varu}) contains an {\it hidden}
symplectic structure. In fact a careful analysis shows that \cite{JS4}:

\begin{equation}\label{sympl}
\Omega_{ij} = \int_{\Sigma_{t}} \Omega ^{\mu} _{ij} \hat{t}_{\mu },
\end{equation}
with,
\begin{equation}\label{dsym}
\Omega ^{\mu }_{ij}:=\frac{\partial E_{j}}{\partial \partial_{\mu }
\varphi^{i}}  
\end{equation}
is antisymmetric (in $i$ and $j$), closed, covariant and conserved
and thus naturally defines a symplectic two-form for {\it first order
theories}. 
Moreover, with these definitions, the basic equations (\ref{gauco}) and
(\ref{varu}) can be 
rewritten on-shell as \cite{JS4}

\begin{equation}\label{brack}
\delta  _{\xi } \varphi^{i} = \left[Q_{_{\xi}}, \varphi^{i}
\right]_{\Omega},
\end{equation}
with,
\begin{equation}\label{syym}
\left[\ ,\  \right]_{\Omega
} := \Omega^{ij}\frac{\delta }{\delta \varphi^{i}} \frac{\delta }{\delta
\varphi^{j}}.
\end{equation}

We then recovered 
the basic Hamiltonian equation (\ref{brack})
which defines the conserved charge in term of the symmetry considered.

\section{The example of gravity}\label{gravex}

The formula (\ref{varu}) can be used for any gauge symmetry. The
examples of Yang-Mills, p-forms, Chern-Simons in D=2n+1 
dimensions, supergravities (in first order formalisms \cite{JS2}) are
treated in \cite{Si,HJS,SiT}.

The purpose of this proceeding is to report in more detail on the
case of gravity for 
two very different boundary conditions, namely spatial infinity
\cite{JS3} ($r=\infty$) and
a Brane-World boundary \cite{Si2} ($r=0$); see figure \ref{spacetime}.

The starting point i)  of the method given in the previous section
is to use a first order formulation of gravity. Two well-known
possibilities can be found in the literature:

$\bullet$ the Palatani formulation, where the metric $g_{\mu \nu}$ and the torsionless
connection $\Gamma^{\rho}_{\mu \nu }=\Gamma^{\rho}_{\nu \mu }$ are
treated as independent fields;

$\bullet$ the Cartan-Weyl formulation, with a vielbein $e^{a}_{\mu }$
and a $so (1,D-1, \Rea)$ connection $_{\perp}\omega_{\mu \ b}^{\ a}$ being the
independent fields.

In the papers \cite{JS1,JS3}, we worked with a third possibility,
namely the Affine gravity which combines both formulations in a nice
way. This theory contains three fields: a metric $g^{ab}$, a vielbein
 $\theta^{a}_{\mu}$ (called canonical one-form) and a $gl (D, \Rea)$
connection\footnote{This connection is assumed neither
torsionless nor metric compatible. Both conditions come from its
own equations of motion, in the so-called {\it Einstein gauge}
\cite{JS1}.} $\omega_{\mu \ b}^{\ a}$ . 
The Palatini formalism is recovered after fixing all the internal $gl (D,
\Rea)$ gauge symmetry with $\theta^{a}_{\mu}=\delta ^{a}_{\mu}$ (the
$\delta$-Kronecker symbol). On the other hand, the vielbein theory (in
its first order form) follows after fixing the internal metric by
$g^{ab}=\eta ^{ab}$ (the flat Minkowski metric). This last choice
breaks $GL (D,\Rea)$ down to $SO (1,D-1,\Rea)$, as illustrated in
figure \ref{affine}.


\smallskip
\FIGURE{\epsfig{file=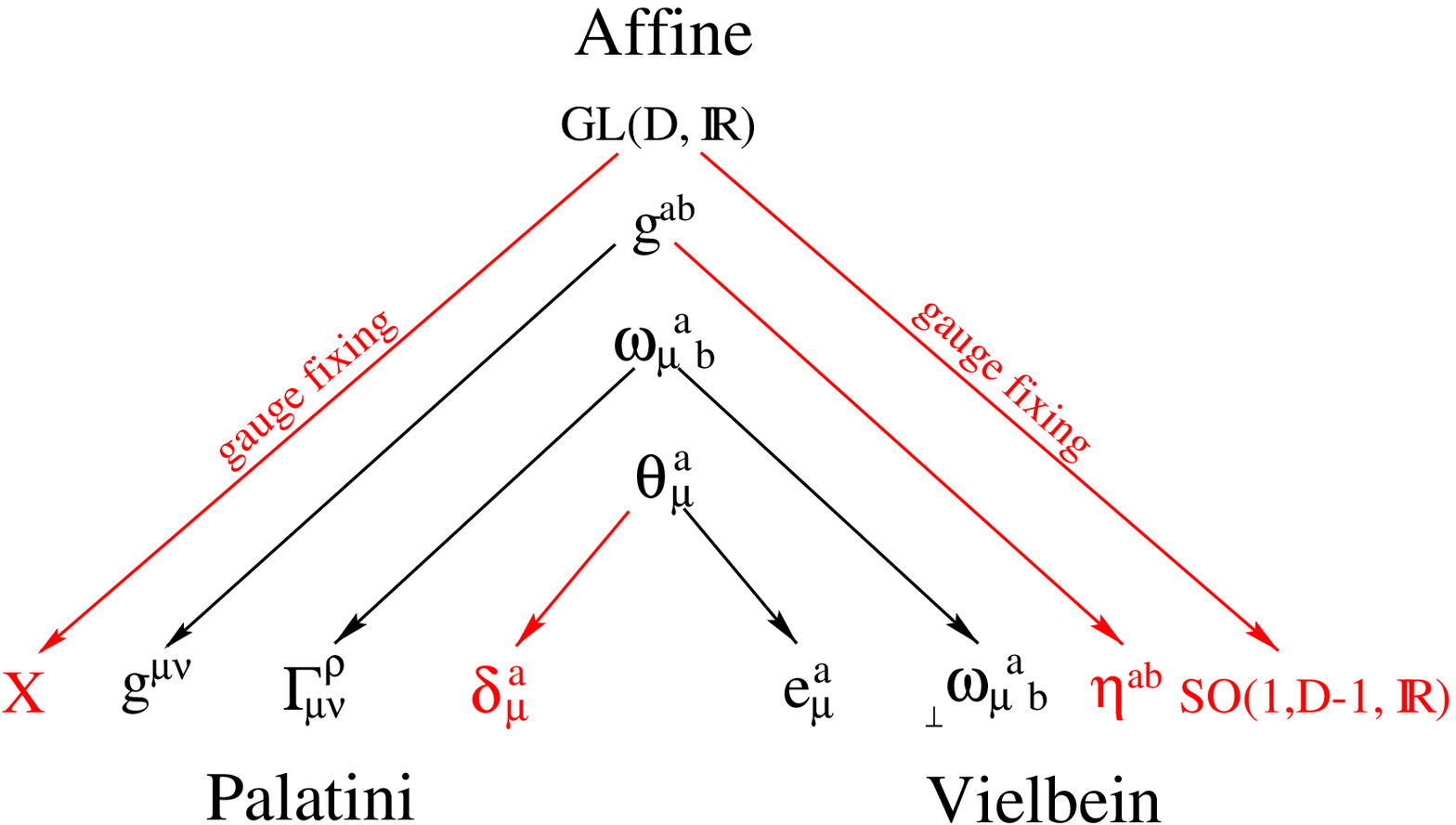,width=7.3cm}%
        \caption{The Palatini and Vielbein formalisms as special cases
of Affine gravity.}%
	\label{affine}}

Using this Affine gravity (needed also for
some technical reasons), we can compute the variation of the
superpotential using equation (\ref{varu}). After going down to the
Palatini formulation following the left arrows of figure \ref{affine}, we
find that the variation of the gravitational superpotential
satisfies\footnote{Do not confuse the arbitrary variation $\delta \varphi^{i}$
with the Kronecker symbol $\delta^{\nu}_{\rho}$.}
\cite{JS3}: 

\begin{eqnarray}
16\pi G\  \delta U_{\xi}^{\mu \nu }
&=& 2\  \nabla_{\rho } \xi
^{\sigma} \delta \left( \sqrt{\left|
g\right|} g^{\rho [\mu} \delta^{\nu ]}_{\sigma } \right) \nonumber\\
& & + 6 \ \delta \left( \Gamma^{\sigma }_{\rho \lambda}\right) \sqrt{\left|
g\right|} g^{\rho [\mu}  \delta^{\nu }_{\sigma} 
 \xi^{\lambda ]} \ \ \  \ \ \label{delu}
\end{eqnarray}

The equation (\ref{delu}) is valid for {\it any} boundary condition
of Einstein gravity (with or without cosmological constant). That
means, it can be used at {\it any} ${\cal B}_{r}$. 
We discuss the cases of spatial infinity and of a Brane-World (namely
$r=\infty$ and $r=0$, see figure \ref{spacetime}) in the
following two subsections. 

\subsection{``Integration'' on ${\cal B}_{\infty}$: The ADM charges}\label{adm}

The boundary conditions at spatial infinity are basically Dirichlet
boundary conditions on the metric,

\begin{equation}\label{dbc}
g_{\mu \nu } \stackrel{r \rightarrow \infty }{\longrightarrow}
\bar{g}_{\mu \nu }, 
\end{equation}
for some given (fixed) asymptotic metric $\bar{g}_{\mu \nu }$. The two usual
cases are the asymptotically flat one, $\bar{g}_{\mu \nu }=\eta_{\mu
\nu }$ and the asymptotically anti-de Sitter one, $\bar{g}_{\mu \nu }=g_{\mu
\nu }^{adS}$.

Using then the boundary conditions (\ref{dbc}), we can ``integrate''
equation (\ref{delu}). The final result \cite{JS3} is a $D$-dimensional version
of the superpotential proposed
by Katz, Bi\u{c}{\'a}k and Lynden-Bell \cite{KBL} as a covariant generalization
of the ADM charges:

\begin{eqnarray}
\ ^{\mbox{\tiny \it
KBL}}U_{\xi}^{\mu \nu } &=& \frac{\sqrt{\left|g \right|}}{16 \pi G}
\left( \nabla^{\mu} \xi^{\nu} - \bar{\nabla}^{\mu}
\xi^{\nu} \right.\label{incomp}\\
& & \left. + (\Delta \Gamma ^{\mu }_{\rho \sigma} g^{\rho
\sigma}-\Delta \Gamma ^{\sigma }_{\rho \sigma }g^{\mu \rho}) \xi^{\nu}
\right) - \mu \leftrightarrow \nu \nonumber
\end{eqnarray}
with,
\begin{equation}\label{gdel}
\Delta \Gamma ^{\mu }_{\rho \sigma}:=\Gamma ^{\mu }_{\rho \sigma}-\bar{\Gamma} ^{\mu }_{\rho \sigma},
\end{equation}
and where the over-bared quantities are computed with the
asymptotic bared metric $\bar{g}_{\mu \nu }$ defined through equation
(\ref{dbc}). 

Note also that the result (\ref{incomp}) is obtained after fixing the
``integration'' constant by requiring:

\begin{equation}\label{erq}
\ ^{\mbox{\tiny \it
KBL}}U_{\xi}^{\mu \nu } (\bar{g}_{\mu \nu }) = 0
\end{equation}
This condition is nothing but a normalization (or definition) of the
classical zero point energy.

As in the Hamiltonian formalism, the diffeomorphism parameter
$\xi^{\rho} (x)$ should be compatible with the Dirichlet boundary 
conditions (\ref{dbc}):

\begin{equation}\label{compt}
g_{\mu \nu } + {\cal L}_{\xi} g_{\mu \nu }\stackrel{r \rightarrow \infty }{\longrightarrow}
\bar{g}_{\mu \nu }.
\end{equation}

The number of non-vanishing $\xi$'s at ${\cal B}_{\infty}$ which
satisfy this equation (\ref{compt}) gives the number of conserved charges. For
instance, an infinite number of them exists 
in $3$-dimensional anti-de Sitter gravity (that is when
$\bar{g}=g^{adS_{3}}$) \cite{BH}.

In summary,
using the general method recalled in section \ref{compsup}, we found for 
Dirichlet boundary conditions (\ref{dbc}) the KBL
superpotential. Moreover, this superpotential is the only one which
satisfies the following properties:

$\bullet$ It is generally covariant.

$\bullet$ If the chosen coordinates are the Cartesian ones of an
asymptotically flat (or anti-de Sitter) spacetime, the KBL
superpotential reproduces the ADM charge formulas (or the
AD \cite{AD} ones for {\it adS}).

$\bullet$ It gives the mass and angular momentum (and the conformal
Brown-Henneaux \cite{BH} charges for {\it adS}$_{3}$ \cite{SiT}) with the right
normalization in any $D\geq 3$.

$\bullet$ It can be used at null infinity and reproduces the Bondi
charges \cite{KL}.

\subsection{``Integration'' on ${\cal B}_{0}$: The Brane-World charges}\label{bwc}

Let us start with the following  
Brane-World action:

\begin{eqnarray}\label{bwac}
{\cal S} &=& \int_{{\cal  M}} \frac{\sqrt{\left|g \right|}}{16\pi G} R
+ \int_{\partial {\cal M}} \frac{\sqrt{\left| \tilde{g} \right|}}{16\pi G}
{\cal K} \nonumber\\ 
& & + \int_{brane} {\cal L}_{brane} (\tilde{g}, \varphi)
\end{eqnarray}
where ${\cal K}$ denotes the extrinsic curvature, $\tilde{g}$ the
pulled-back metric on the brane and $\varphi$ some Brane-World fields.

In more general cases (basically for supergravity
applications), one or several scalar fields are included in the
bulk, together with some D-dimensional cosmological constant. 
However, it can be checked that the final result, namely equation
(\ref{fres}), 
remains {\it unchanged} under these modifications
\cite{Si2}. That is why we consider here the simplest action
(\ref{bwac}) for illustrative purposes.

The presence of the Gibbons and Hawking \cite{GH} boundary term ${\cal K}$ in
the action (\ref{bwac}) can be explained by the following. The 
variational principle, namely $\delta {\cal S}= 0$ (on-shell),
implies the following boundary condition {\it on the brane}:

\begin{equation}\label{junct}
-\frac{\sqrt{\left|g \right|}}{16\pi G} {\cal  K}_{\mu
\nu}= T_{\mu \nu }^{bra} -\frac{1}{D-2} \tilde{g}_{\mu \nu } T^{bra}
\end{equation}
with
\begin{equation}\label{energ}
T_{\mu \nu }^{bra}:=\frac{\delta {\cal L}_{bra}}{\delta
\tilde{g}^{\mu \nu }}, \ \ T^{bra}:=\tilde{g}^{\mu \nu } T_{\mu \nu }^{bra}.
\end{equation}

The equation (\ref{junct}) is nothing but 
the Lanczos-Israel \cite{LI}
junction condition. The link between this condition (\ref{junct}) and
the extrinsic curvature boundary term  
in (\ref{bwac}) was first pointed out in the context of four-dimensional
gravity\footnote{In the work of \cite{HL}, the ``Brane-World'' was
just an infinitely thin shell of dust.} in \cite{HL} (for more recent
work, see \cite{Gl}). For related works from the Brane-World scenario
point of view, see \cite{CR,DM,BV}.

The boundary conditions (\ref{junct}) now fix the normal derivative of
the metric. These are then Neumann conditions, in opposition to the
previous Dirichlet case (\ref{dbc}). Following the method of section
\ref{compsup},
we can again ``integrate'' the basic equation (\ref{delu}), using now
the boundary conditions (\ref{junct}).  After several simplifications,
the final 
expression for Brane-World charges (\ref{gauco}) is extremely simple
\cite{Si2}:
\begin{equation}\label{fres}
 \ ^{\mbox{\tiny BW}}Q_{\xi }  
=  2 \int_{B_{0}}  \hat{T}^{bra}_{\mu \nu } t^{\mu }
\xi^{\nu}
\end{equation}
where $\hat{T}^{bra}_{\mu \nu }$ is ``almost'' the energy-momentum
tensor of the brane Lagrangian (compare with (\ref{energ})):
\begin{eqnarray}
\hat{T}^{bra}_{\mu \nu } &=& T^{bra}_{\mu \nu }+\frac{1}{2}
\tilde{g}_{\mu \nu } {\cal L}^{bra} \nonumber\\
&=& \sqrt{\left| \tilde{g} \right|} \frac{\delta }{\delta
\tilde{g}^{\mu \nu }} \left(\frac{{\cal L}_{bra}}{\sqrt{\left|\tilde{g} \right|}} \right).\label{almos}
\end{eqnarray}

Note that it should be possible to recover the result (\ref{fres})
using the Hamiltonian formalism together with the Regge and Teitelboim
procedure \cite{RT}.

The differences between the BW charges (\ref{fres}) (with
(\ref{almos})) and the usual energy-momentum tensor in flat
spacetime are quite important:

$\bullet$ $ \ ^{\mbox{\tiny
BW}}Q_{\xi }$ gives the {\it total} energy (for $\xi^{\mu}=t^{\mu}$),
including the 
gravitational contribution. Therefore, if our real world is a
Brane-World, the gravitational energy can be {\it localized}, with a
covariant local density given by (\ref{almos}).

$\bullet$ $ \ ^{\mbox{\tiny
BW}}Q_{\xi }$ vanishes if the Brane-Lagrangian depends on the
pulled-back metric only through an overall $\sqrt{\left|\tilde{g}
\right|}$ factor (see the examples below).

$\bullet$ The positivity of total energy would require a modified
energy condition, namely:
\begin{equation}\label{mode}
\hat{T}_{\mu \nu}^{bra} t^{\mu } \xi^{\nu} \geq 0,\  \mbox{\it for } \xi ^{\nu
}\  \mbox{\it timelike.} 
\end{equation}

\smallskip

We will finish with a simple example where the formula (\ref{fres})
can be tested. Let us consider the action (\ref{bwac}) together with a
bulk scalar field:
\begin{equation}\label{biulk}
-\int_{\cal M}\sqrt{\left|g \right|}\left( \frac{1}{2}g^{\mu \nu}
\partial_{\mu } \phi \partial_{\nu} \phi + V (\phi)\right).
\end{equation}

As we commented after equation (\ref{bwac}), the result (\ref{fres})
is unchanged by the addition of the action (\ref{biulk}). This is
proven in detail in \cite{Si2}.

We will study the simplest example where the Brane-Lagrangian is a
cosmological constant,

\begin{equation}\label{branl}
\int_{brane} \frac{\sqrt{\left|\tilde{g} \right|}}{2} \lambda (\phi),
\end{equation}
together with the ans{\"a}tze,

\begin{eqnarray}
ds^{2} &=& e^{2 A (r)} \eta_{ij} dx^{i} dx^{j} + dr^{2}\label{ans}\\
\phi &=& \phi (r) \label{anjj} 
\end{eqnarray}
for the bulk metric and the scalar field. This simple model is studied in
detail in several papers, see for instance \cite{ST,DFGK,Ka}. 

Using now (\ref{branl}) in our result (\ref{fres}), we immediately
find that the Brane-World charges vanish:

\begin{equation}\label{van}
\ ^{BW}Q_{\xi} = 0.
\end{equation}

In particular, the energy is zero. There is another independent way to
recover this result. In fact, in the static case (\ref{ans}), we
expect to find an agreement between the energy density and minus the total
Lagrangian. This is {\it indeed} the case when the Gibbons-Hawking
boundary term 
${\cal K}$ of (\ref{bwac}) is properly taken into
account. Using the equations (\ref{branl}) and (\ref{ans}-\ref{anjj}) in
(\ref{bwac}) and (\ref{biulk}), we find:

\begin{eqnarray}
{\cal H} = -{\cal L} &=& e^{4 A}\left(5 A'^{2}+2 A''+\frac{1}{2} \phi'^{2}
+ V  \right)\nonumber\\
& & -2 \partial_{r} \left( e^{4 A} A'\right) -\frac{1}{2}
\partial_{r}\left(  e^{4 A} \lambda  \right), \ \ \label{enez}
\end{eqnarray}
where the prime (as $\partial_{r}$) denotes a differentiation with
respect to $r$. 

Note that the first term in the second line of (\ref{enez}) comes from
the Gibbons-Hawking boundary term and 
was
not considered in \cite{ST,DFGK,Ka}.
It is then straightforward to check that ${\cal H}$ indeed 
vanishes using the equations of motion and the boundary conditions, in
agreement with (\ref{van}) (see again \cite{Si2} for details). 

A similar conclusion arises for the Brane-World de Sitter ansatz:

\begin{equation}\label{desitt}
g_{ij}^{dS} :  -dt^{2}+e^{2\sqrt{\Lambda } t}
(dx_{1}^{2}+dx_{2}^{2}+dx_{3}^{2})
\end{equation}

The five-dimensional metric is then given by (\ref{ans}),
together with the replacement $\eta _{ij}\rightarrow g_{ij}^{dS}$.
The Brane-World Lagrangian is again (\ref{branl}) and so the result (\ref{van})
remains unchanged. As for the previous example, the vanishing of the
energy can be checked in 
an independent way: The ansatz (\ref{desitt}) is not anymore
static and then the Hamiltonian is now given by ${\cal H} = \pi_{\mu \nu
}\dot{g}^{\mu \nu } - {\cal L}$, with $\pi _{\mu \nu }$ the canonical
momenta conjugated to the metric. Using the equations of
motion and the boundary conditions, it is straightforward to check
that ${\cal H}$ indeed vanishes, again in agreement with (\ref{van}).

Note finally that the vanishing of energy in the supersymmetric
extensions  was
pointed out in \cite{BKP}.

\bigskip

\noindent {\bf \large Acknowledgments}

I would like to thank the organizers and in particular B. Julia for
nice discussions and collaboration in part of this work. 
I am also grateful to the EU TMR contract FMRX-CT96-0012 for financial support.

\end{document}